\documentclass[prb,epsf,epsfig]{revtex4}
\usepackage{amsmath}
\usepackage[toc,page]{appendix}
\usepackage{graphicx,pdflscape}

\newcommand{\beq}{\begin{equation}}
\newcommand{\eeq}{\end{equation}}
\newcommand{\barray}{\begin{eqnarray}}
\newcommand{\earray}{\end{eqnarray}}
\newcommand{\lab}[1]{\label{#1}}
\newcommand{\disp}[1]{Eq.~(\ref{#1})}
\newcommand{\refdisp}[1]{Ref.~(\onlinecite{#1})}
\newcommand{\figdisp}[1]{Fig.~(\ref{#1})}
\newcommand{\nn}{\nonumber}

\begin{document}
\title{Nature of Split Hubbard Bands at Low Densities   }
\author{ Daniel Hansen, Edward Perepelitsky and B Sriram Shastry }
\affiliation{Physics Department, University of California,  Santa Cruz, Ca 95064 }
\date{\today}
\begin{abstract}

We present a numerical scheme for the Hubbard model that throws light on the rather esoteric nature of the   Upper and Lower Hubbard bands that have been invoked often in literature. We present a self consistent solution of the ladder diagram equations for the Hubbard model, and show that these provide, at least in the limit of low densities of particles, a vivid picture of the Hubbard split bands. We also address the currently topical problem of decay of the doublon states that are measured in optical trap studies, using the ladder scheme and also by an exact two particle calculation of a relevant Greens function.
\end{abstract}
\maketitle

\section{Motivation and Introduction\lab{sec1} }

Hubbard's introduction of split bands in Ref.[\onlinecite{hubbard1}], i.e. the so called upper Hubbard band (UHB) and the lower Hubbard band (LHB), is  one of the most important qualitative ideas in the theory of correlated electrons.  Their origin is the idea that since the energy levels of the atomic limit show two sets of states, one at $\omega \sim 0$ and another at $\omega \sim U$ as in \disp{split-bands} below, the formation of a crystal would broaden these levels into two sets of sub-bands.  These sub-bands were originally discussed by Hubbard using  a non perturbative technique, that has the advantage of being exact in the limit of vanishing bandwidth $W\to0$, i.e. the atomic limit.
However, the technique  failed to produce a Fermi liquid for weak couplings, as one expects physically. This failure  led to severe early criticism of Hubbard's work\cite{herring}.  The problem of reconciling Fermi liquids with the local picture developed by Hubbard, leading to the split bands, is of great importance in  the  physics of strong correlations. The one exception is the dynamical mean field theory that gives a good account of the sub-band formation, especially in the proximity of half filling\cite{vollhardt,georges}. However, away from half filling, the picture is obscure and remains largely unresolved.
It is this task that we address in the present work. We study the ladder diagrams that are argued to be exact at low densities, sharpen the argument for their validity in terms of the self energy, and show that at least in this limit, the concept of the split bands is completely consistent with the Fermi liquid picture. The numerical solution of the ladder diagrams is carried out in a self consistent way and shows the emergence of the Hubbard split bands for large enough $U/W$.  These merge for weak couplings and our results give a vivid picture of   the  crossover   from weak to intermediate to strong coupling. 

The self energy is momentum and also frequency dependent in the ladder scheme, and  for low densities
 provides a full picture of the renormalization processes that occur at arbitrarily large interaction scale $U$. In particular we see that the spectral function shows a low lying feature and a high energy $\sim O(U)$ feature, with spectral weights that are equal  to $1- \frac{n}{2}$ and $ \frac{n}{2}$ respectively. It is seen that every single added particle thus depletes the weight of the LHB and adds to the UHB, thereby accomplishing  a ``long range spectral transfer''- that has been described in literature as ``Mottness''\cite{phillips,sawatzky}.

The momentum space occupancy $m(k) = \langle c^\dagger_{k \sigma} c_{k \sigma} \rangle$
 is computed and it is usefully broken up into three parts \disp{mk}.  The occupied part $m_{1}(k)$ in \disp{mk}, corresponding to  occupied states that are automatically inside the LHB, the unoccupied LHB part $m_{2}(k)$ corresponding to unoccupied LHB states, and the unoccupied UHB part $m_{3}(k)$. In the limit of $U \to \infty$,  only $m_{1}$ and $m_{2}$ survive, and this projection gives an exact view of the physics of the $t$-$J$ model as well in the low density limit. At low densities we find  that the ladder diagrams lead to a Luttinger Ward compliant Fermi surface, and this Fermi surface survives the limit $U \to \infty$. Thus even in this limit of extreme correlations $U \to \infty$, adiabatic continuity to the Fermi gas  holds. Therefore  we have a useful and  concrete  alternative to the  extreme coupling ideas proposed in work by one us \cite{ECQL}, where a different Fermi volume emerges at {\em all densities}, including the lowest ones.

One contemporary context for the Hubbard split bands is the problem of high $T_{c}$ superconductors, here Anderson\cite{anderson-rvb} has eloquently argued that for large $U$, one can confine attention to carriers in the LHB, with the UHB pushed out of the range of relevant states. Given  this projection to the LHB, the charge carriers  inherit exotic properties such as spin charge separation, and also a new interaction, namely the super exchange that comes with a scale of $t^2/U$. We see that at least at low densities where the ladder scheme is valid, the LHB does separate out cleanly for $U\geq W$, but the carriers are  yet subject to Fermi liquid behaviour.

Another recent context for motivating this work is the study of the Hubbard model far away from equilibrium  with cold atom realization\cite{esslinger,demler}, where the carriers in the UHB are optically excited, and their lifetime studied by measuring the overlap of the excited state with the initial state. We find that a calculation of a related correlation function is possible in the Fermi liquid at low densities, albeit in a close to equilibrium situation unlike the experiments. We are also able to exhibit the correlation function exactly for a pair of particles in the Hubbard band. Interestingly, the resulting life times show some similarity in functional dependence  to those found in experiment, although with a very different time scale.

\section{Ladder Scheme Equations at Low Density \lab{sec2} }

The ladder scheme for the Greens function Refs.[\onlinecite{galitskii,kanamori,horowitz,fukuyama,randeria}] corresponds to
convoluting a  particle-particle ladder scattering amplitude $\Gamma(Q)$ with a single Greens function $G(k)$ to form the self energy $\Sigma(k)$ as follows: 
\barray
\Gamma(Q)& = &\frac{U}{1+U \Pi(Q)}, \nn \\
\Pi(Q)& = & \frac{1}{\beta N_s}\sum_{p}G(p)G(Q-p),  \nn \\
\Sigma(k)& =&\frac{1}{\beta N_s}\sum_{p}G(p)\Gamma(k+p). \lab{eq1}
\earray
Here $N_{s}, N_{e}$ are the number of sites and electrons, $n= N_{e}/N_{s}$ is the electron number density, and we use the  notation $k=(\vec{k}, i \omega_{k})$ with imaginary odd frequencies
$\omega_{k}=  \pi \frac{1}{\beta} ( 2 k+1)$  of the finite temperature field theory\cite{agd} for Fermions, and 
reserve the capital letters for Bosonic frequencies, e.g. $Q= (\vec{Q}, i\Omega_{\nu})$  and $\Omega_{\nu}=    2 \pi \frac{1}{\beta} \nu$. Here the summation over $p$ represents a sum over the vector component and also the imaginary frequency. A paramagnetic state is assumed and the spin label is suppressed for brevity. In addition to \disp{eq1}, we have the  Dyson equation $G^{-1}(k)=G_{0}^{-1}(k) - \Sigma(k) $ with the usual non interacting Greens function $G_{0}^{-1}(k) = i \omega_{k} - \varepsilon_{k}+ \mu$. 
Thus  the ladder scheme is a self consistent non linear scheme that needs to be solved numerically for the various objects $G(k), \  \Sigma(k), \  \Pi(K)$.  We can solve for the Dyson equation in the ladder scheme iteratively:
 \barray
G^{-1}(k)&=&G_{0}^{-1}(k)-\frac{1}{\beta N_s}\sum_{p}G(p)\frac{U}{1+\frac{U }{\beta N_s}\sum_{q}G(q)G(k+p-q)}. \lab{dyson}
\earray
 For example in the first step 
 we can calculate the scattering amplitude (and self energy) using $G_0$ and use Dyson's eqn to obtain a new Green's function we call $G_1$:
\barray
G_{1}^{-1}(k)&=&G_{0}^{-1}(k)-\frac{1}{\beta N_s}\sum_{p}G_0(p)\frac{U}{1+\frac{U }{\beta N_s}\sum_{q}G_0(q)G_0(k+p-q)}. \lab{first}
\earray
We may continue and compute $G_{2}(k)$ using $G_{1}(k)$ to recompute the self energy (i.e. the second term in \disp{first}), and repeat this process iteratively to obtain  $G(k)= \lim_{n \to \infty }G_{n}(k)$.  The difference between $G_{1}(k)$ and the fully self consistent $G(k)$  arises from the repeated renormalizations implicit in the full equations, and this brings about the self consistent broadening of several sharp features that arise in $G_{1}(k)$. In Fig.~(\ref{spec2d}) we discuss the difference in the spectral functions from these two theories as an illustration of this phenomenon.

  Alternatively we start by  introducing spectral representations for the various quantities of physical interest\cite{agd,mahan}:
\barray
G(\vec{k}, i \omega_{k})&=& \int \ d \nu \ \frac{\rho_{G}(\vec{k},\nu)}{i \omega_{k}- \nu}, \nn \\
\Sigma(\vec{k}, i \omega_{k})&=& U \frac{n}{2}+  \int \ d \nu  \ \frac{\rho_{\Sigma}(\vec{k},\nu)}{i \omega_{k}- \nu} ,\nn \\
\Gamma(\vec{Q}, i \Omega_{Q})&=& U+  \int \ d \nu  \ \frac{\rho_{\Gamma}(\vec{Q},\nu)}{i \Omega_{Q}- \nu}. \lab{spectral}
\earray
The spectral functions $\rho_{\Gamma}(\vec{Q},\nu)$ etc have a compact support  and are therefore convenient for numerical integration on a suitably discretized grid of frequencies. The numerical solution is performed after using a spectral representation for various physical quantities. We first turn the Dyson equation \disp{dyson} into 
a non linear integral equation for the spectral function from \disp{spectral} as follows:
\barray
\rho_{\Sigma}(\vec{k} ,\omega) & = &\frac{1}{N_s}\sum_{\vec{p}}\int d\nu  \ \rho_{G}(p,\nu) \  \rho_{\Gamma}(\vec{p}+\vec{k} ,\nu+\omega) \  (f(\omega)+n_B(\omega+\nu)), \nn \\
\rho_{\Pi}(\vec{Q},\Omega)) & = & \sum_{\vec{q}}\int d\nu \ \rho_{G}(q,\nu) \ \rho_{G}(Q-q,\Omega-\nu) \ (f(\nu)+f(\Omega-\nu)-1), \nn \\
\rho_{\Gamma}(\vec{Q},\Omega)&=&\frac{-U^{2} \rho_{\Pi}(\vec{Q},\Omega)}{(1+U Re \ \Pi (\vec{Q},\Omega))^{2}+(\pi U \rho_{\Pi}(\vec{Q},\Omega))^{2}}, 
\lab{spectral-eqs}
\earray
with $f(\omega)$ and $n_{B}(\omega)$ as the Fermi and Bose distribution functions $[\exp{ \beta \omega} \pm 1]^{-1}$, and $Re \ \Pi( \vec{Q},\Omega)$ defined as the  Hilbert transform of $\rho_{\Pi}(\vec{Q},\nu)$, i.e. 
$$ Re \ \Pi( \vec{Q},\Omega)= {\cal P} \int \ d \nu \ \frac{\rho_{\Pi}(\vec{Q},\nu)}{\Omega-\nu}.$$

\subsection{Low density limit and self energy sum rule }

The original  argument for the ladder scheme\cite{kanamori,galitskii} is that it is exact in the low density limit. This argument is borrowed from the theory of nuclear matter, where Brueckner\cite{brueckner} originally argued that 
at any order $n$ of perturbation theory for the ground state energy (i.e. the Goldstone diagrams), the dominant diagrams are those with  the smallest number of downward lines of holes.  Topologically there need to be at least two  such hole lines in the free energy diagrams.  The particle particle ladder diagrams have only two hole lines at any order. Thus the ladder diagrams dominate all others at each order in perturbation theory.  Importantly for nuclear matter, this logic shows that the large (divergent) two body interaction is not a problem, it is cut off by these ladders, giving  in the end an expansion in a dimensionless parameter obtained by combining the two body scattering length with the average inter particle separation. A parallel argument for bosons was provided  by Lee,  Huang and Yang \cite{yang}.  The Kanamori- Galitskii papers implement  this   idea for the Feynman diagrams, where one has additionally hole hole scattering, in addition to particle particle ladders- for structural reasons that distinguish the Ferynman diagrams from the Goldstone ones.
 However these extra terms do not detract from the particle particle ladders that cohabit the Feynman series and provide a particular $O(n^2)$ correction term.

The reader would note that the above argument is rather  indirect, in particular it gives us  no clue to why we should accept the self energy that emerges from this scheme as exact. In this context, it is useful to note that the self energy satisfies an exact  series of sum rules\cite{harrislange,sawatzky,hess}, of which the lowest is 
\beq
s_0(k) \equiv \int \ d \nu \ \rho_{\Sigma}(\vec{k},\nu) = U^{2} \ \times \frac{ n(2-n)}{4}, \lab{sumrule}
\eeq
where the RHS is {\em independent of $\vec{k}$}.  Note that this sum rule is valid for arbitrarily large $U$ and at all densities. We can use this as a check of our calculation by testing for the $\vec{k}$ independence of the computed LHS, and also monitor its weight relative to the RHS. The self consistent solution of the ladder diagrams contain the low density limit and also provide some uncontrolled results at higher densities, and it is important to know the limit on density to which we can trust these results. 
\begin{figure}
\includegraphics[width=3.5in]{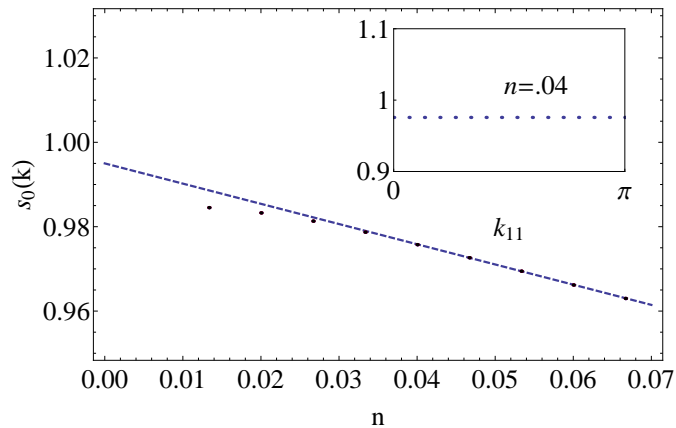}
\caption{The zeroth moment of the self energy versus the  density normalized to the exact value $  U^{2} \ \times \frac{ n(2-n)}{4}$. This data is in 2-dimensions with U=10,W=2. The inset shows the $k$ independence of the sum rule along the (11) direction for the case n=.04, with  variations in the sixth significant figure.}
\label{zerothmom}
\end{figure}
Fig.~(\ref{zerothmom}) gives details of this test for the ladder diagrams.  For higher densities  the ladder diagram theory is systematically   wrong for the $O(n^2)$ term,  since we can show analytically that  at large $U$ and low density $s_0(k)= U^2 \times \frac{n (1-n)}{2}+O(U)$ in contrast to \disp{sumrule}.

\subsection{The Atomic Limit \lab{atomic}}
\begin{figure}
\includegraphics[width=5.5in]{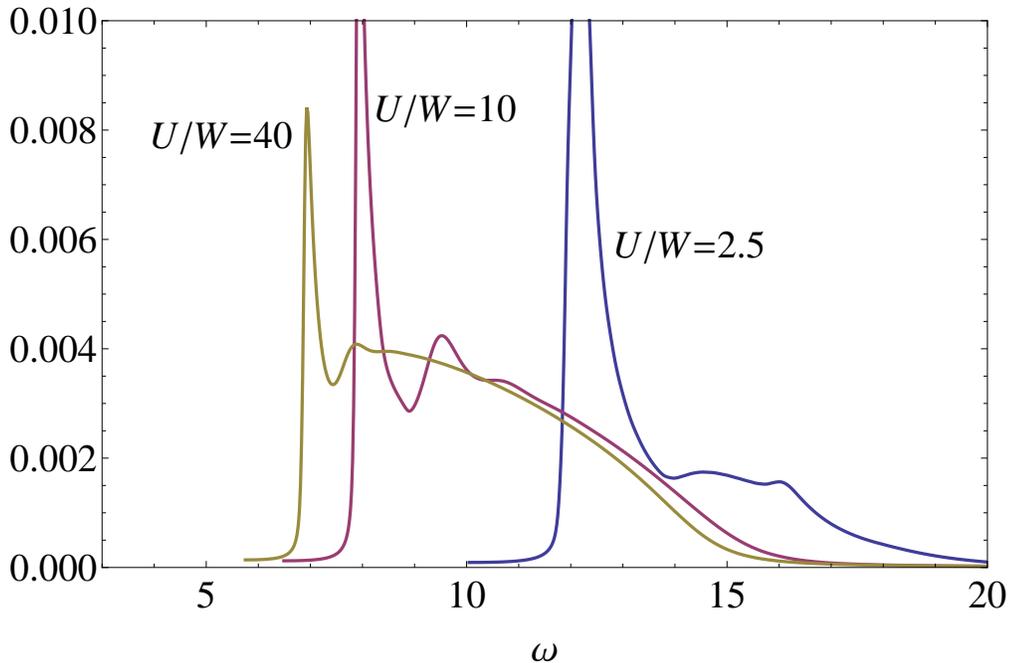}
\caption{High frequency (UHB)  DOS in 2D, U=10, n=1/20;  As the hopping is decreased, the UHB feature does not become narrow,  but rather  maintains a width of O(U). The sharp k-dependent features narrow as the hopping decreases. The broad continuum is essentially k-independent, showing very little dependence on the hopping in the limit of strong coupling.  In this limit, the UHB becomes completely independent of the bandstructure. In figure \ref{spec2d} the broad UHB of the full band can be seen with the Hubbard-1-like $G_1$ superimposed. In $G_1$, the UHB feature is broadened only by $\eta$ and the LHB is suppressed for clarity.}
\label{atomlim2d}
\end{figure}

We discuss briefly the atomic limit, i.e. a limit where $U$ remains finite but the band width $W\to0$, this is the 
limit where one can solve for the Greens function exactly quite simply. 
\barray
\Sigma_{Atomic}&=&U \times \frac{n}{2}+U^{2} \times \frac{\frac{n}{2}(1-\frac{n}{2})}{i\omega+\mu-U(1-\frac{n}{2})} \\
G_{Atomic}&=&\frac{1-\frac{n}{2}}{i\omega+\mu}+\frac{\frac{n}{2}}{i\omega+\mu-U} \label{split-bands}
\earray
The breakup of the Greens function into two parts, with energies $\sim 0$ or $\sim U$  and weights $1-n/1$ and $n/2$ is of course the  fundamental factor that leads one to the picture of upper and lower Hubbard bands. Hubbard's contribution\cite{hubbard1} was to provide a Greens function for finite hopping $W$ using an equation of motion method that extended the Atomic limit, although the details of his treatment came in for severe criticism \cite{herring} due to the failure of his scheme to ever yield a Fermi liquid with the Luttinger Ward\cite{luttinger_ward} ordained Fermi surface. The present scheme of ladder diagrams achieves this interpolation smoothly and exactly, if only in the limit of low densities. From \figdisp{atomlim2d}, we see that the sharp feature is accompanied by a broad background of width $O(U)$  that presumably arises from the uncontrolled $O(n^2)$ corrections to the ladder diagram self energy sum rule \disp{sumrule}.

\subsection{Emergence and structure of the Split bands of Hubbard \lab{sec3}}

\begin{figure}
\includegraphics[width=5.5in]{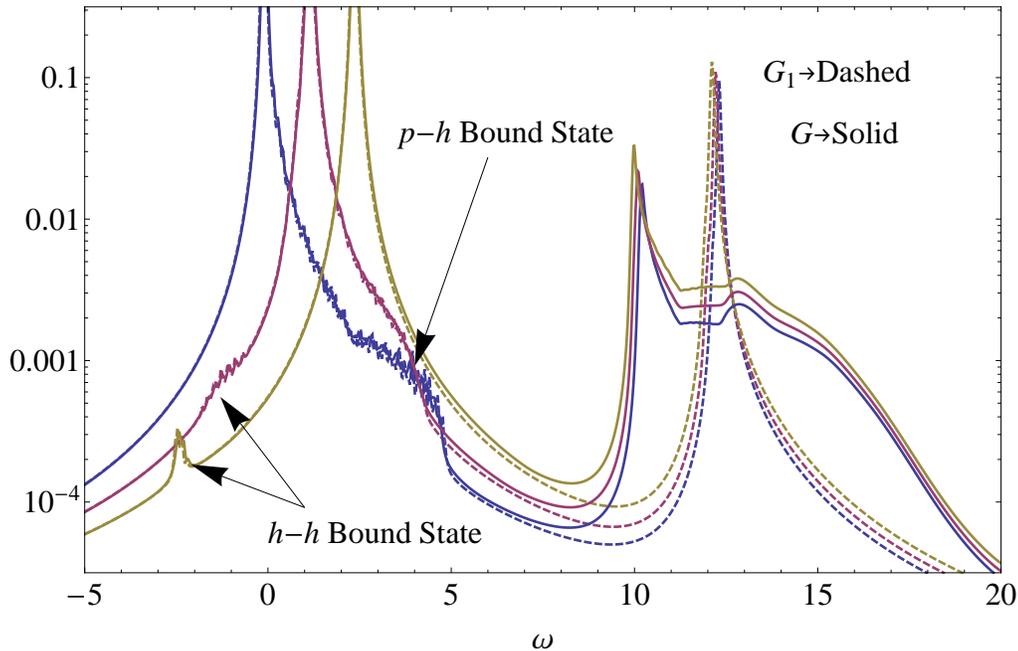}
\caption{2D, U=10,W=2.5, n=1/20;  The spectral function at three values of the wave vector $(0,0), \ (\frac{\pi}{2},\frac{\pi}{2}), \ (\pi,\pi)$, in blue, red and gold colours. 
Besides the quasiparticles we observe three features emerging in each spectral function. Most obvious is the UHB feature which lies at a $\omega \approx O(U)$ and integrates to a weight of $n/2+O(n^2)$. This feature is dramatically broadened in the self consistent G also becoming less k-dependent. On each edge of the quasiparticle band we observe small dispersing features. Ref.(\onlinecite{randeria}) have previously identified the negative frequency feature as a 2-hole antibound state while Ref.(\onlinecite{anderson}) has discussed a particle-hole antibound state just above the quasiparticle band. These features are essentially unchanged in going from $G_1$ to the exact G. \label{spec2d}}
\end{figure}

In the ladder diagrams, it is straightforward to identify
the origin of the upper Hubbard band:  the scattering amplitude $\Gamma(Q)$ at frequencies $\Omega_{Q}\sim U$, has a pole in the first iteration, i.e. at the level of $G_{1}$ with 
\beq
\Gamma_{1}(Q) \equiv \Gamma(Q ; [G_{0}] ) \sim \frac{U^{2} (1-n)}{ i \Omega - U (1-n)} . \lab{gutzpole}
\eeq
This pole was noted very early in works \refdisp{horowitz,shastry-rice} who identified this pole as the origin of strong correlations and Gutzwiller type factors.  In Fig.~(\ref{spec2d}), we see that the spectral function obtained from the first iteration i.e. $G_{1}$ shows a sharp feature at a higher energy of $O(U)$ that arises from this pole.
This peak disperses and may be viewed as a ``baby version'' of the upper Hubbard band.
 Next a self consistent treatment of this theory with $\Gamma(Q;[G])$ evaluated with $G$ (rather than $G_{0}$) broadens the upper band substantially as seen in Fig.~(\ref{spec2d}). It is interesting that the lowe Hubbard band, i.e. the structure at energies below $U$ are stable with respect to the iterations, and are hardly different between the first iteration scheme and the final one.
 
 We also see in Fig.~(\ref{spec2d}), the existence of two features that have been commented upon in literature.
 The feature near the band bottom that disperses, is the so called hole-hole bound state note by Randeria and Englebrecht \refdisp{randeria}, whereas the hump near the leading edge is a particle hole bound state feature noted by Anderson \refdisp{anderson}. These features  coexist with the other, dominant ones, namely the quasiparticle peak of the Fermi liquid and the broadened upper Hubbard band peak. If we replace the log linear scale in Fig.~(\ref{spec2d}) with a linear linear scale as in Fig.~(\ref{3func}), the UHB becomes almost negligible compared to the LHB feature.

\subsection{ Frequency Dependent Self energy }
We next display the self energy in the ladder scheme. The spectral density for the self energy is given in \disp{spectral-eqs}, and it is possible to obtain an equation for its momentum sum, i.e. a local self energy density
\beq
\frac{1}{N_s}\sum_{k}\rho_{\Sigma}(k,\omega)=\int d\nu \rho_{G,loc}(\nu) \rho_{\Gamma,loc}(\nu+\omega) (f(\omega)+n_B(\omega+\nu)).
\eeq
For comparison, we note that the local self energy in the atomic limit considered in Section \ref{atomic} is given by
a single delta function centered at $ U(1-\frac{n}{2})-\mu $ as:
\beq
\rho_{Atomic}(\omega) = U^{2} \ \frac{n}{2} ( 1- \frac{n}{2} ) \ \delta[\omega+\mu - U(1-\frac{n}{2})].
\eeq

\begin{figure}
\includegraphics[width=5.5in]{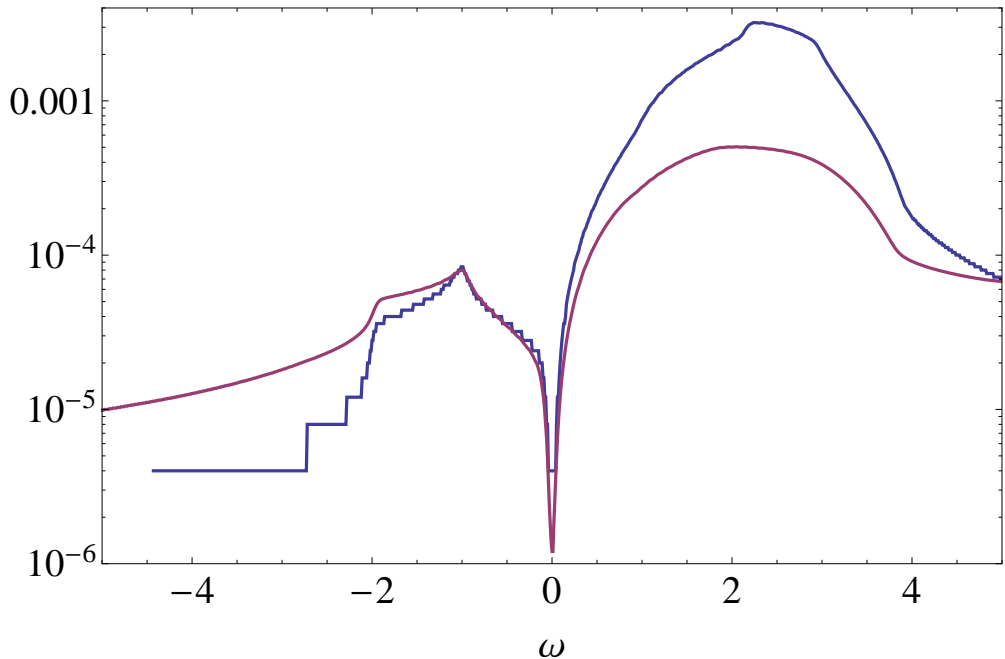}
\caption{2D, \ U=.25, \  U=10, W=2, n=.049. The  local $\rho_{\Sigma}(\omega)$ divided by U versus $\omega$. 
The two  chosen values of U  are in the weak coupling (blue $U=.25$ ) and strong coupling (red $U=10$) ranges respectively. We see at the lowest temperatures that the the self energy  curves overlap when scaled by U displaying a characteristic quadratic dip at the chemical potential.  \label{ggg}}
\end{figure}

We also note the form of this object for a Fermi liquid at finite $T$ 
\beq
\rho_{Local}^{Fermi Liquid}(\omega) =  a \ \omega^{2} + f_{Background}(\omega),
\eeq
a simple second order self consistent theory (corresponding to truncating the ladders at the first rung) gives the picture of this in a Fermi liquid Fig.~(\ref{ggg}).

\begin{figure}
\includegraphics[width=5.5in]{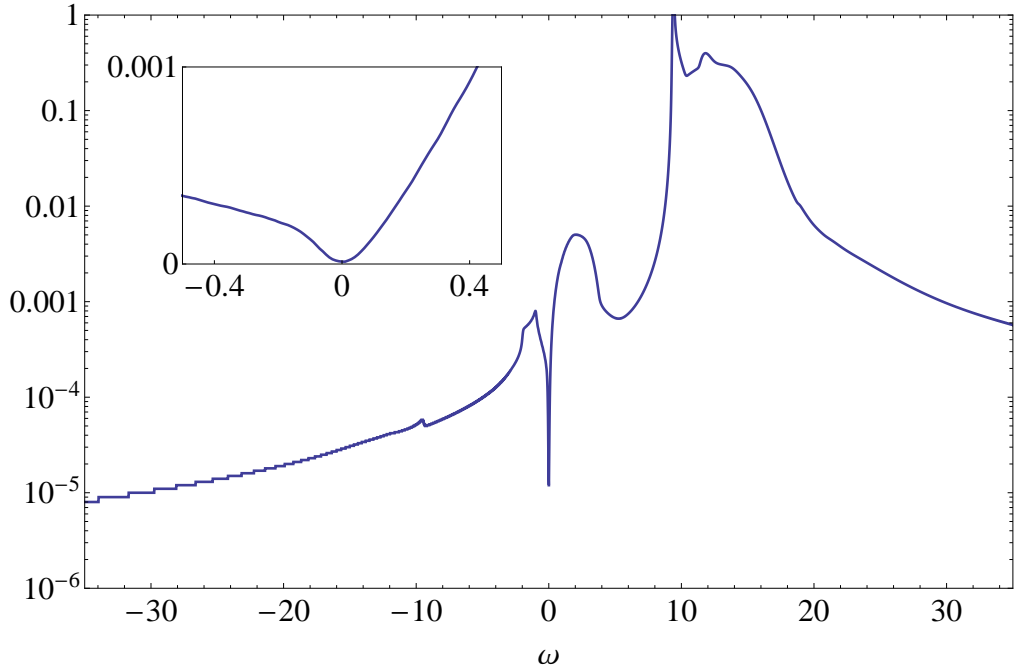}
\caption{The local self energy spectrum in 2D, U=10, W=2, n=.05 The log scale plot shows the full scale of the UHB. The inset highlights the quadratic minimum at low energies. The quadratic minimum drops below the scale of $\eta$ so it can be said to represent an infinite lifetime. }
\label{lowfreq}
\end{figure}

\begin{figure}
\includegraphics[width=5.5in]{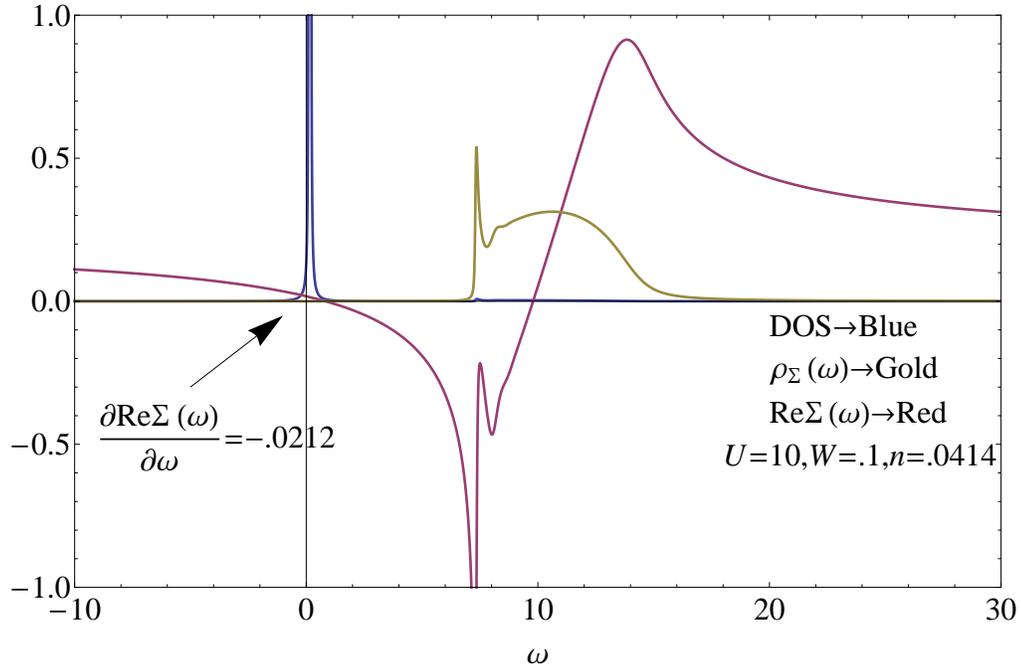}
\caption{The 2D  DOS i.e. the momentum averaged  spectral function $\rho_{G}(\vec{k},\nu)$  and the momentum averaged $\rho_{\Sigma}(\vec{k},\nu) $.  The LHB feature is the sharp peak near $\omega \sim 0$. The UHB feature in the DOS is nearly invisible here but lies just below the feature in $\rho_{\Sigma}$ scaled down by a factor of $\omega^2$. The real part of the self energy  for $\omega \geq 0$ initially drops linearly with frequency over a  range $W \ll \omega \sim \frac{U}{2}$, as required in the limit of extreme correlations \cite{ECQL,ECQL-linslope}. It then flips at the threshold of the UHB,  rising  across the range of the UHB until at the highest energy it begins to decay down towards the Hartree term at infinite energy. }
\label{3func}
\end{figure}

We see in Fig. \ref{lowfreq} that the ladder scheme inherits both a quadratic minimum at $\omega =0$ from the Fermi liquid and a large and broad feature near $\omega \sim U$ from the emergent Hubbard upper band. The inset emphasizes the Fermi liquid aspect, and the reader will observe that the absolute scale of this function is dominated by the UHB feature. In Fig. \ref{3func} the density of states of the Greens function $\rho_{G}(\vec{k},\nu)$ is illustrated, along with the real and imaginary parts of the self energy. The small feature in the DOS at the energy scale $U$ is the UHB.  We see that the real and imaginary parts of the self energy reflect its presence in a profound fashion, that would be hard to guess from the size of the peak. In detail, it is interesting that the real part of the self energy does display a linear behaviour in $\omega$ with a known slope as one expects in the intermediate frequency range $0 \ll \omega \ll U$ from the theory of extremely correlated electronic systems in Ref. ~(\onlinecite{ECQL,ECQL-linslope}).

\begin{figure}
\includegraphics[width=4.5in]{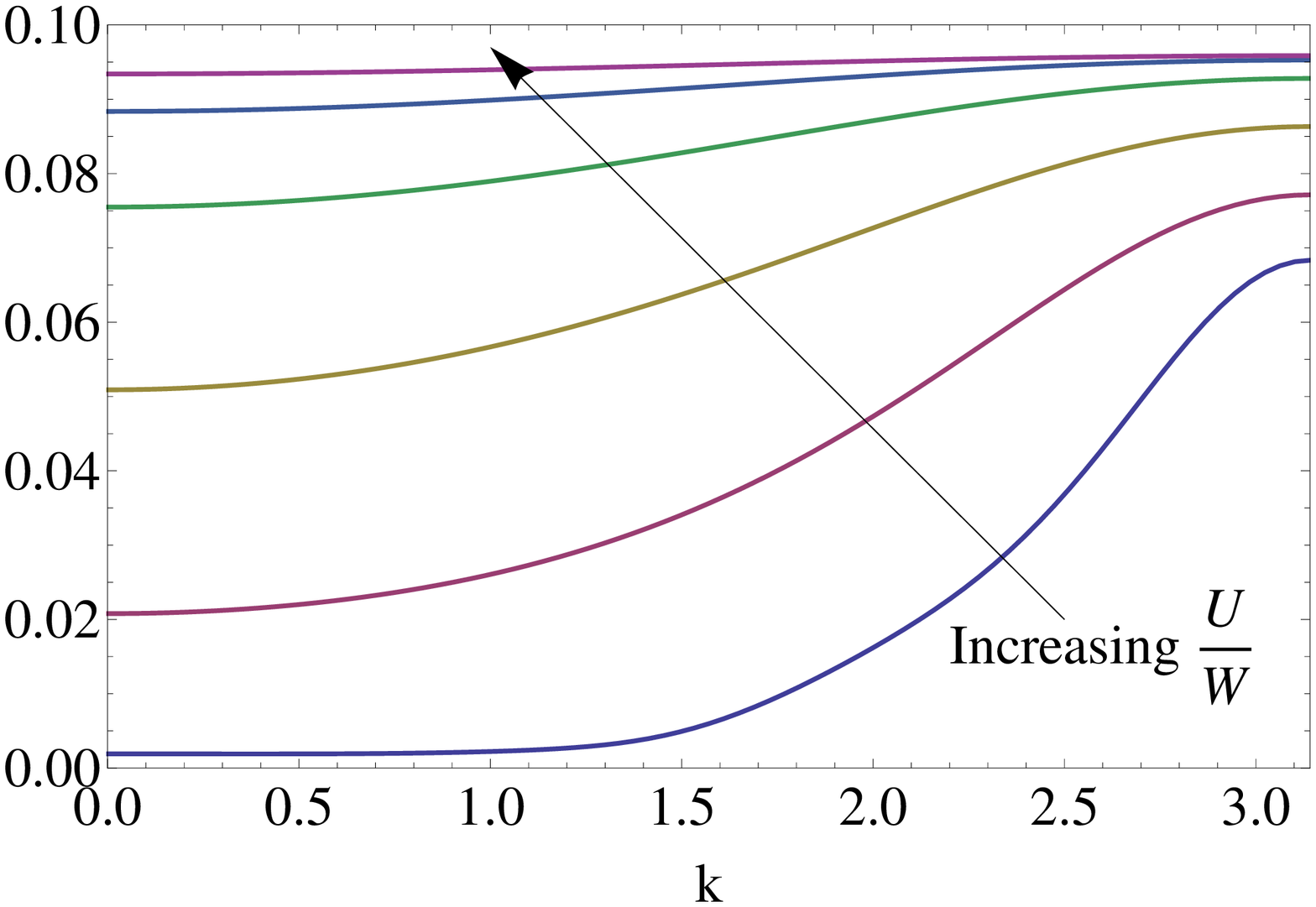}
\caption{The integrated spectral weight over the UHB is called $m_3(k)$. It is plotted here for $\frac{W}{U}=1.6,.56,.196,.0686,.024,.0085$.  In this case n=.15.  We observe that the weight of the UHB  exceeds n/2 and becomes flat as U/W tends to infinity. }
\label{kUHBw}
\end{figure}

\begin{figure}
\includegraphics[width=4.5in]{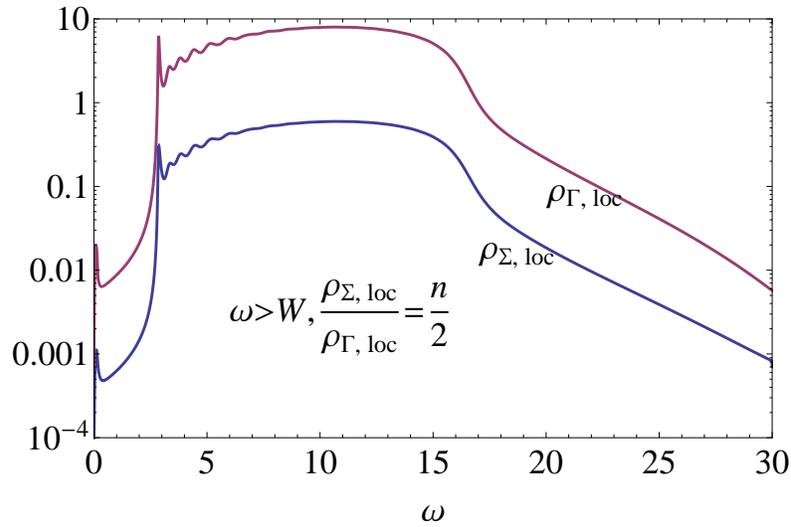}
\caption{From the convolution structure of $\rho_{\Sigma}(k,\omega)$ we see that the local objects of $\Sigma$ and $\Gamma$ are related by the ratio $n/2$ when $\omega>W$ for all values of $W/U$. In the strong coupling limit where the upper band is essentially independent of k, this relationship will be approximately true for each wavevector. On the negative frequency side, the thermal function act differently such that the ratio for $\omega<-W$ is approximately $(1+\frac{n}{2})$.}
\label{rhorho}
\end{figure}

When $W=0$ the UHB has a weight which is independent of momentum. However, for finite  $W$,  momenta  near the top of the band will transfer weight more readily to the UHB.
\figdisp{kUHBw} illustrates this progression.
We show in \figdisp{rhorho} that the behaviour of the local spectral function $\langle \rho_{\Gamma}(\vec{Q}, \nu)\rangle_{Q }$ closely follows that of the local self energy $\rho_{\Sigma}(\nu)$.
If we look at large $\omega$ such that we can make the approximation $\omega+\nu \approx \omega$, the integral for $\rho_{\Sigma}(k,\omega) $ in \disp{spectral-eqs} reduces to
\beq
\frac{1}{N_s}\sum_{k}\rho_{\Sigma}(k,\omega)\sim \frac{n}{2} \ \rho_{\Gamma,loc}(\omega),
\eeq
accounting for the similarity of these in  Fig. \ref{rhorho}.

\subsection{Momentum occupancy}
\begin{figure}
\includegraphics[width=5.5in]{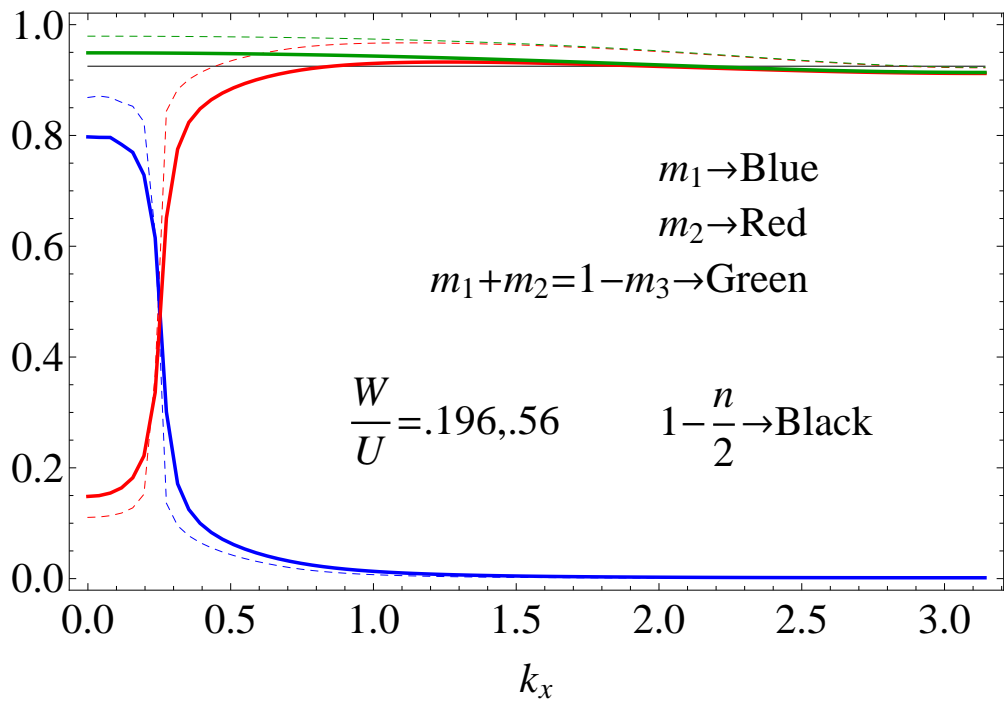}
\caption{1D U=10 W=.56 (dashed), W=.196 (solid), n=.15 1d T=.005. Here  $m_1$ is essentially the zero temperature quasiparticle occupation, while $m_2$ accounts for the LHB particle addition spectrum, The sharp step in occupation occurs precisely at the Luttinger Fermi surface which satisfies the Luttinger Ward sum rule. The sum of $m_1$ and $m_2$ is less than one due to the weight transferred to the upper band. The total lower band weight approaches 1-n/2 as U/W goes to infinity. }
\label{m1m2}
\end{figure}

We next turn to the momentum occupancy $m_{k}=  \langle c^{\dagger}(k)c(k) \rangle$;  this can be obtained from the Greens function or $\rho_{G}(k,\nu)$ by integration over the frequencies. In order to understand and illustrate  the nature of the LHB and UHB breakup of this important object, we carry out the integration up to the Hubbard-Mott gap energy $\omega_{g}$. This energy  scale is well defined  when $W \ll U$, and in case of smaller $U \sim W$ it requires a definition.  In our work, it is  operationally defined as the energy where the spectral density   $\langle \rho_{G}(k,\nu)\rangle_{k}$ is minimum. Thus we define three objects $m_{j}(k)$ with $j=1,2,3$ 
\barray
m_1(k)&=&\int_{-\infty} ^{0} d\omega \rho_G(k,\omega)\\
m_2(k)&=&\int_{0}^{\omega_{g}} d\omega \rho_G(k,\omega)\\
m_3(k)&=&\int_{\omega_g}^{\infty} d\omega \rho_G(k,\omega). \label{mk}
\earray
Here $m_{1}(k)$ represents the momentum space occupancy of the occupied states that lie below the chemical potential. These are automatically in the LHB for energetic reasons, and satisfy the sum rule $\sum_{k}m_{1}(k)= n/2 \times N_{s }$ with a sum over the  entire Brillouin zone (BZ). Next $m_{2}(k)$ represents the LHB contribution to the unoccupied states, since the chemical potential lies within the LHB.  If we send $U \to \infty$ then we are left with only the LHB, and in that limit, we expect the sum 
$m_{1}(k)+ m_{2}(k)= 1- \frac{n}{2}$ pointwise at each $k$. However for finite but large  $U$ this sum differs from  $ 1- \frac{n}{2}$ by terms of $O(t/U)$, and the UHB comes into play. Indeed $m_{3}(k)$ refers to precisely the UHB contribution to the momentum occupation, and its momentum average over the BZ is $\frac{n}{2}$. These are displayed for typical parameters in \figdisp{m1m2}.
The sum of all three m functions should add to unity for each wave vector. However, due to the finite frequency resolution of our numerics this sumrule is only approximately satisfied. We limit the error to $<1\%$ by reducing our frequency step $d\omega$. The error is concentrated near $k_f$ where the spectral function is sharpest.

In \figdisp{m1m2}, we display the $k$ dependence of the three occupancy functions for a typical set of parameters. It is clear that the Luttinger Ward Fermi surface controls the variations of the functions $m_{1}$ and $m_{2}$, which complement each other so that the sum is almost a constant.

\section{Doublons and their dynamics \lab{sec4}}
\subsection{Doublon Decay in the low density limit}
In the recent experiments\cite{esslinger,demler} the lifetime of doublons created by optical excitation of the trapped atoms has been carried out, providing us with an added impetus for this study. The experiments  actually study the decay of  a highly non equilibrium initial state  $| \psi_{Initial}\rangle $ with {\em a finite fraction of excited doublons}, i.e. $\langle \psi_{Initial} |\hat{D} |\psi_{Initial}\rangle \propto N_{s}$, where the doublon number $\hat{D} = \sum_{i }n_{i \uparrow}n_{i \downarrow}$.  The object studied is the time evolution of such a state followed by a measurement of $D$ and then a projection on to the evolved state i.e.
\beq
\xi(t_{r})=  \langle \psi_{Initial} | \ \exp{ \{ i t_{r} H \} } \ \hat{D} \ \exp{\{ - i t_{r} H \} } \ |\psi_{Initial}\rangle. \lab{xi}
\eeq
Here and below  we use the symbol $t_{r}$ to denote real (Schr\"odinger) time, thus distinguishing it from the band hopping parameter $t$.
Such a correlation function is not usually amenable to  study near equilibrium type situations studied in many body physics.  The initial state is itself quite far from being an equilibrium (ground) state. However, in the limit of very low densities, one can approximately view the initial state as the vaccuum or few particle state with a  few doublon excitations- and within this picture we may ask  how a single doublon decays. This is roughly the question of the lifetime of a state in the upper Hubbard band, and thus related to our general theme  in this work.

We are able to calculate the lifetime of a doublon within the ladder scheme, and hence presumably an exact answer  at low densities as argued here. 
We next provide a discussion of the function $\gamma$ in a low density Fermi liquid. We start with the correlation function defined for Matsubara time $\tau \geq 0$ in terms of the two particle Greens function\cite{agd}
\beq
\gamma(r,\tau)\equiv G^{II}_{\uparrow,\downarrow,\downarrow,\uparrow}(r \tau ,r \tau;0,0 ) =\langle c_{r,\uparrow}(\tau)c_{r,\downarrow}(\tau)c_{0,\downarrow}^{\dagger}(0)c_{0,\uparrow}^{\dagger}(0)\rangle,
\eeq
and an analogous expression for real times $ \gamma(r,t_r)$. 
This object can be expressed in terms of the scattering amplitude\cite{derivation1} as
\beq
\gamma(r,t_r)=\sum_{Q}\int d\Omega \rho_{\Gamma}(Q,\nu)(1+n_B(\nu))e^{-iQr-i\nu t_r}. \label{gamma-ladder}
\eeq

\begin{figure}
\includegraphics[width=3in]{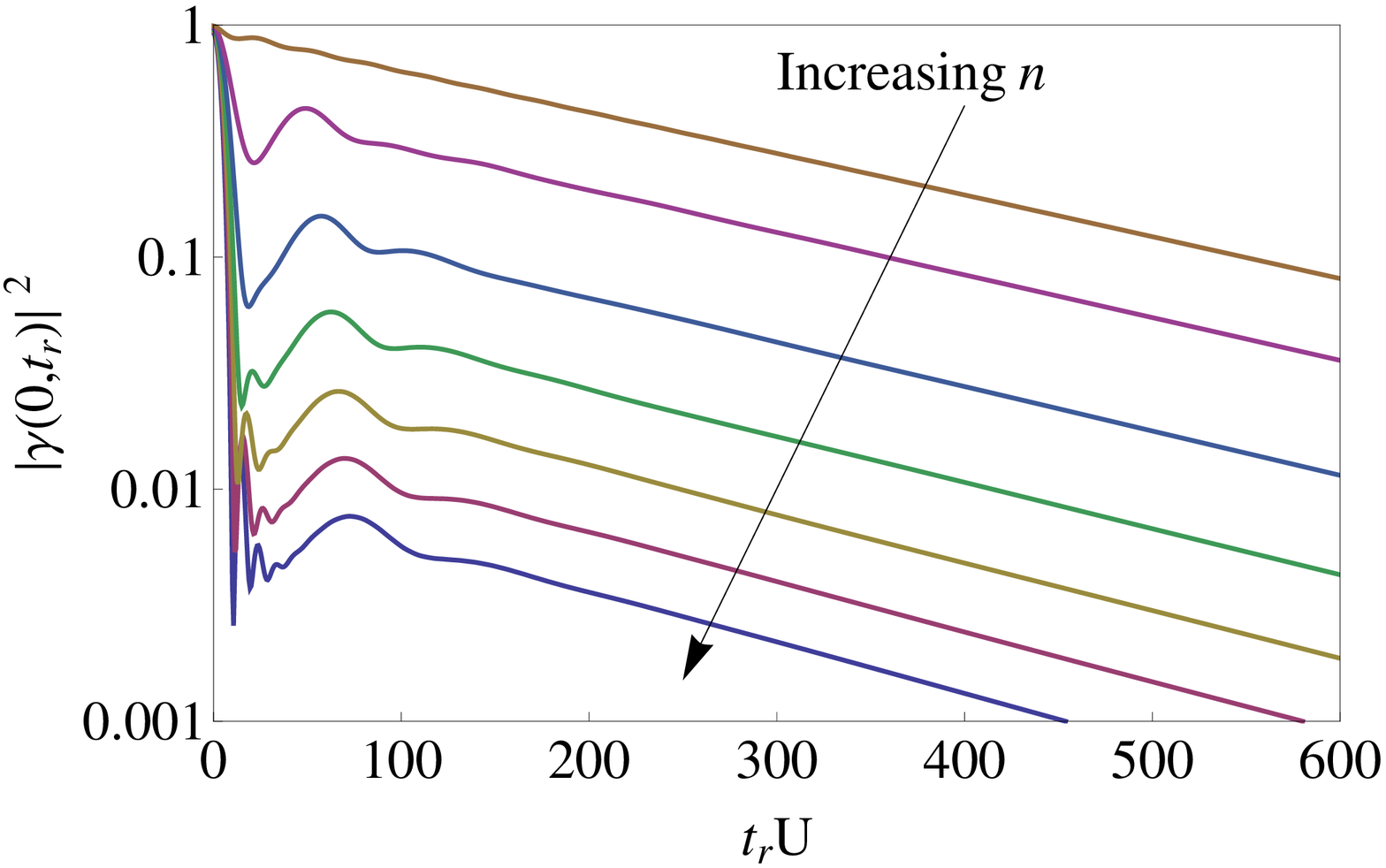}
\includegraphics[width=3in]{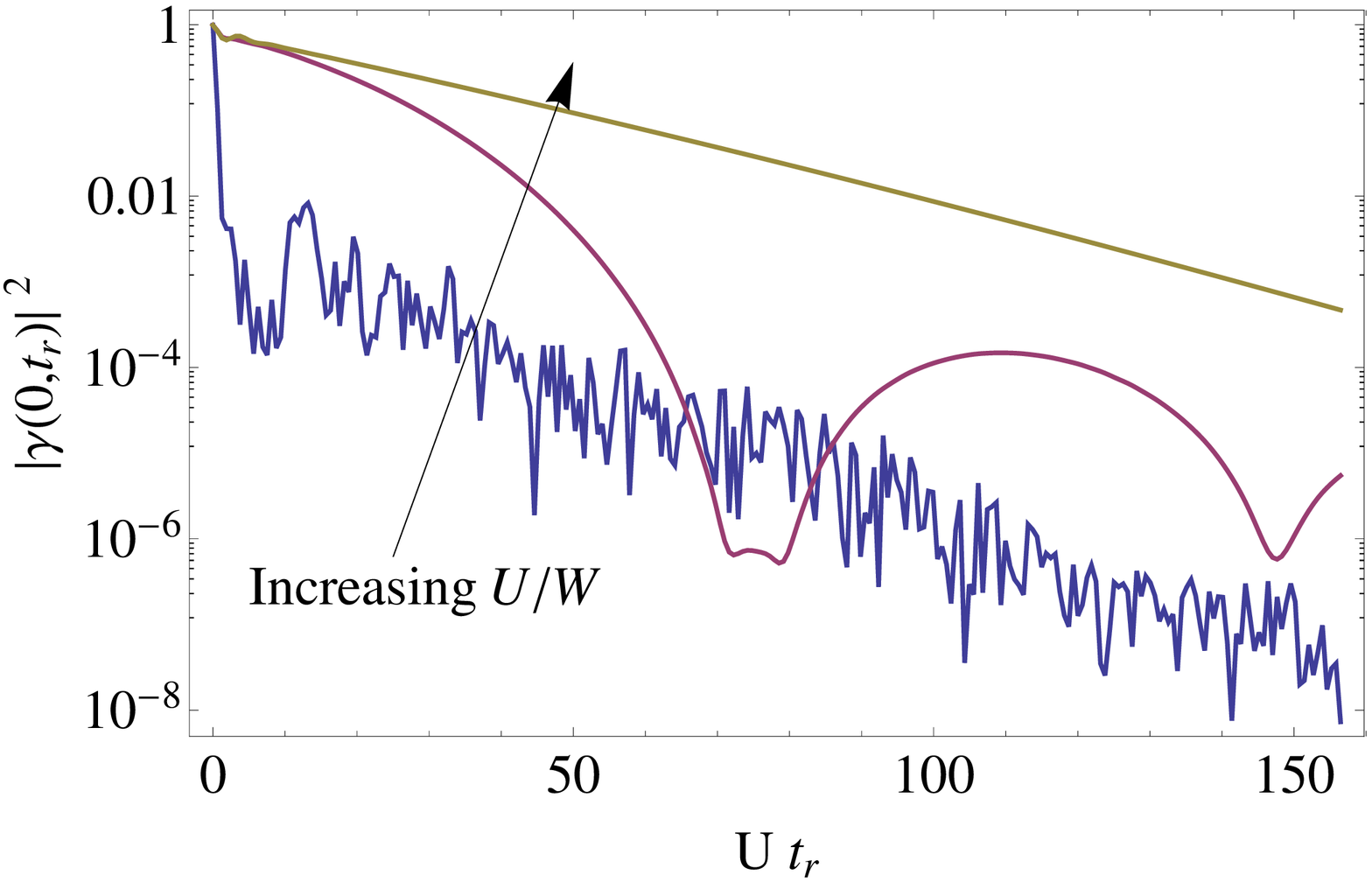}
\caption{The doublon dynamics breaks into two regimes: a sharp decay at early times followed by a long exponential tail. The magnitude of the initial decay depends strongly on the density. In the limit  $n\to 0$ the initial decay disappears,  indicating that the UHB is comprised of sharp features only in the limit of vanishing density.
In the right panel, the $U$ dependence of the long time decay is shown, it slows down and is finally limited by the level broadening $\eta$ assumed in our numerics.}
\label{doubloc2dn}
\end{figure}
In Fig. \ref{doubloc2dn}, we display $\gamma(0,t_r)$ within the ladder scheme. As the density is increased, the UHB becomes broader and less k-dependent, however sharp k-dependent features persist with weight which decreases as n goes to zero. The k-dependent pieces remain sharp and determine the rate of the long time exponential decay. On the other hand  the k-independent pieces, being broad, determine the short time decay. Due to our finite frequency resolution these numerics do not see the long time exponential decay becoming infinitely long once $t<\eta$. 

\begin{figure}
\includegraphics[width=3.5in]{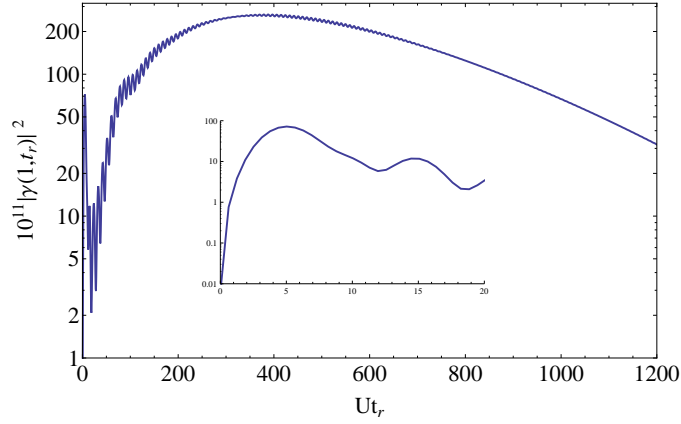}
\caption{   The inset shows  that $\gamma(1,t_r)$ goes to zero at early times since there is no mechanism to hop at small times. On a longer time scale we see the development of an exponential decay. The small magnitude of the correlation is due to fact that the UHB is largely k-independent.   }
\label{offsite}
\end{figure}

We have also computed the off site correlation function $\gamma(1,t_r>0)$, Fig. \ref{offsite} shows that even the site directly adjacent the created doublon has a very  small amplitude.
\subsection{ Exact Solution of the Doublon Decay Problem for Two Particles.}
In addition to the discussion of the low density case, we are able to solve exactly the admittedly simple problem of the dynamics a single doublon in the Hubbard model, and from this study provide some feeling  for the validity of the ladder scheme. The single  doublon problem is solvable since for two particles of opposite spin, we have a total momentum quantum number and in each sector of this, we have a single particle type Schrodinger  
equation to solve.  Let us first outline this problem and its solution with regard to the correlation function
\beq
\gamma(r,t_{r})= \langle0\mid c_{r,\uparrow}(t_r)c_{r,\downarrow}(t_r) c_{0,\downarrow}^{\dagger}(0)c_{0,\uparrow}^{\dagger}(0)\mid0\rangle. \lab{doublon-cf}
\eeq
 Here the average is with respect to the vaccuum state with no particles, although below we will use the average over the thermal distribution function for a low density Fermi liquid. In the case of two particles, it is in fact  possible to show that $\gamma(r,t_{r})$ is related to the correlator $\xi(t_{r})$ in \disp{xi} exactly through
\beq
\xi(t_{r})= \sum_{r} |\gamma(r, t_{r})|^{2} \lab{xi2}. 
\eeq
This follows upon using the fact that with only two particles in the system, the destruction operator $c_{r,\uparrow}(t_{r})c_{r,\downarrow}(t_{r})$ can only connect to the vaccuum state. We expect this relation to be only approximately true for a dense Fermi system but useful since it can be computed with relative ease by one of several techniques. It is also dominated by the term $r=0$ as shown explicitly below in Fig.\ref{offsite}, and hence it is useful to regard $|\gamma(0,t_{r})|^{2}$ as an estimator of $\xi(t_{r})$.

In  \refdisp{demler}, Demler et. al. estimate $\gamma(0,t_{r})$ by an argument that is appropriate in an incoherent Fermi system, and estimate that this function decays on  a time scale that is given as
\beq
\frac{h} {\tau} = A \ t \exp{\{ -  B \ \frac{ U }{W} \}} \lab{decay}. 
\eeq 
The vanishing of the rate as $W\to 0$ is expected in view of the conservation of the doublon number in the absence of electron hopping, the coefficients are estimated from experiments on the 3-d cubic lattice ($W= 12 t$)  as $A\sim .9 \pm 0.5 $, and $B \sim 1.6 \pm 0.16 $.

For the two particle problem, we have exact analytical and numerical solutions. In the interesting case of
$U>W$ in $d$ dimensional hypercubes with nearest neighbor hopping, we can write
\barray
\gamma(0,t_{r})& = & \gamma_{L}(0,t_{r}) + \gamma_{U}(0,t_{r}),  \nn \\
\gamma_{U}(0,t_{r})&\sim&  e^{-i(U+4d\frac{t^2}{U}) t_{r}} J_0^d(\frac{4t^2}{U}t_{r}) ,
\earray 
where the LHB contribution $\gamma_{L}\sim O((W/U)^2)$ and negligible. The second  term  arises from the UHB, and for intermediate $W \ll U$  is related to the Bessel function $J_{0}$ whereby it decays as a power law rather than as an exponential. This is understandable since the two body problem is an integrable system, and we expect that in the low density limit, this power law would be replaced by an exponential type decay.  The function $|\gamma|^{2}$
can be found easily (see Appendix ) by numerical means and Figs.\ref{ddecay} and \ref{ddecay2}  give us a picture of the decay.
\begin{figure}
\includegraphics [width=4.5in]{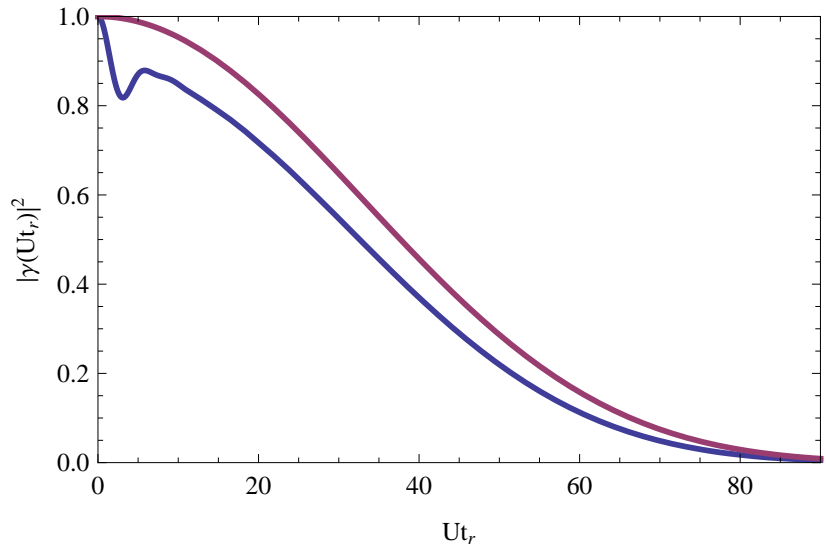}
\caption{Doublon decay on a cubic lattice with $U=15$ and $W=12$. The shape of $\mid \gamma_{U}(Ut_{r})\mid^2$ (red curve) initially deviates slightly from the exact numerical result (blue curve)  due to the neglect of the  $\gamma_L$ term, which decays much more quickly than the UHB contribution.}
\label{ddecay}
\end{figure}
\begin{figure}
\includegraphics [width=4.5in]{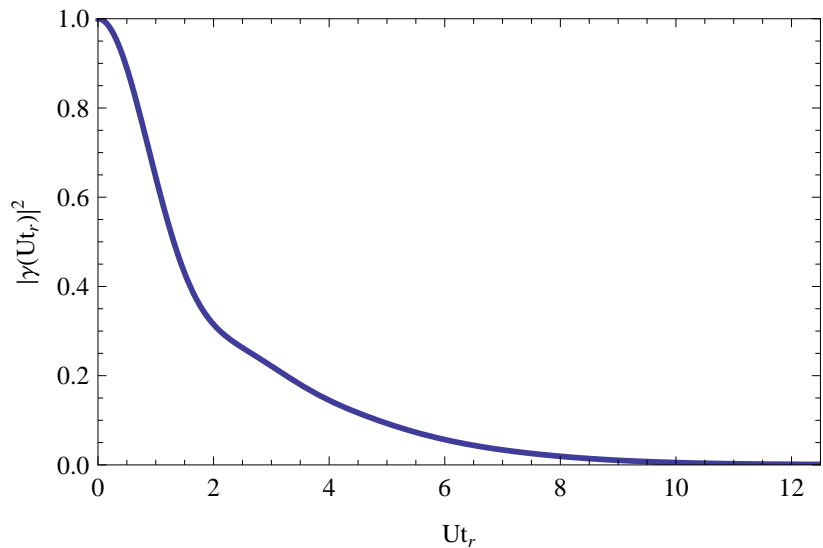}
\caption{Doublon decay on a cubic lattice with $U=5$ and $W=12$. In the case $U<W$, it is much more difficult to find an exact analytical form, so only the numerical result is displayed. }
\label{ddecay2}
\end{figure}
\begin{figure}
\includegraphics [width=4.5in]{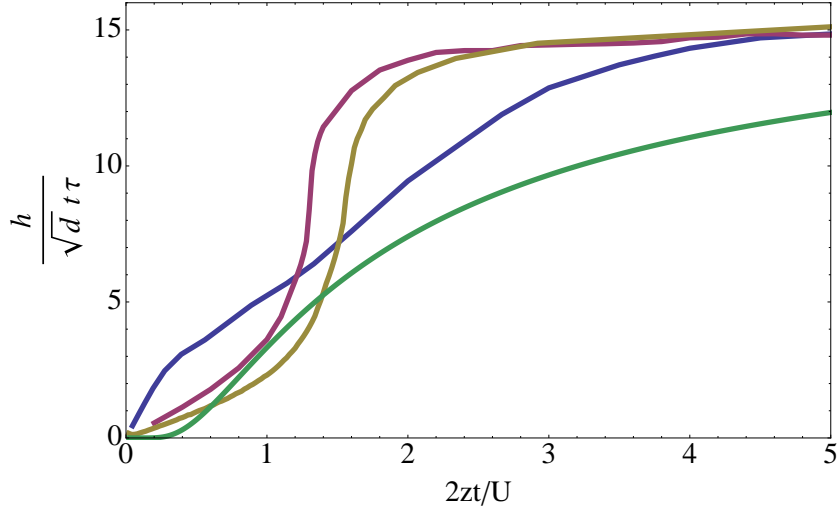}
\caption{Two theoretical calculations, from ladder diagrams of \disp{gamma-ladder} in 2-dimensions (blue)  and the exact 2 particle solution  from \disp{xi2} and  \disp{ex-2} in 2- and 3- dimensions (red and gold). These are  compared to the experiment \disp{decay} in 3-dimensions (green), scaled  to coincide at weak coupling by   a factor $26.4$.  The theory and expermient are in very different limits of physical  parameters,  but have a similar shape except at  large  $U/t$.  
 }
\label{edward2}
\end{figure}

In Fig. \ref{edward2}, we show that the Half Width at Half Max (HWHM) of the computed $\gamma(0,t_{r})$ leads to a rate $\frac{\hbar}{\tau_{HWHM}}$ which  has a behaviour that is similar to that in the experiments \disp{decay}.

\section{conclusions}
In conclusion, we have shown that the self consistently computed ladder diagrams    provide a detailed  picture of the split bands for the Hubbard model. The UHB has a distinct shape that is captured here and related to the shape of the two particle scattering amplitude. We have delineated how the lower Hubbard band occupation is influenced by the passage to large $U$. Here  the background momentum occupance found in variational studies of the Gutzwiller approximation\cite{rice-2} arise here dynamically. Finally, we have shown that the decay of the doublon in such a system can be calculated by the ladder diagrams as well as by exact methods for very low densities, and the shapes of these curves are fairly close to those found in recent experiments on atomic traps performed under very different physical conditions.

\begin{acknowledgements}
This work is supported by DOE through a grant BES-DE-FG02-06ER46319. We are grateful to D. Huse, H. R. Kishnamurthy and M. Rigol for helpful discussions. 
\end{acknowledgements}
\appendix
\section{   Exact correlation functions for  the  two particle Hubbard Model }

 We consider the Hubbard model with two particles, one spin up and the other spin down. Our goal is to calculate the following correlation function.
\begin{equation} \gamma(t_{r}) = \langle 0 \mid c_{i\downarrow} c_{i\uparrow}e^{-iHt_{r}} c_{i\uparrow}^\dagger c_{i\downarrow}^\dagger\mid0\rangle = \gamma_{U}(t_{r}) + \gamma_{L}(t_{r}).
\end{equation}
The two parts arise from intermediate states that are in the two split bands. Thus
\barray 
 \gamma_U(t_{r}) & = &  \sum_{\nu\epsilon UHB}\mid\langle\nu\mid c_{i\uparrow}^\dagger c_{i\downarrow}^\dagger\mid0\rangle\mid^2e^{-iE_\nu t_{r}}  \nn \\
 \gamma_L(t_{r}) &  = &  \sum_{\nu\epsilon LHB}\mid\langle\nu\mid c_{i\uparrow}^\dagger c_{i\downarrow}^\dagger\mid0\rangle\mid^2e^{-iE_\nu t_{r}}.
 \earray
 We now calculate the eigenvalues and eigenstates for the 2 particle Hubbard model. As our basis we take momentum eigenstates.
\begin{equation} \mid Q,k\rangle \equiv c_{Q-k\uparrow}^\dagger c_{k\downarrow}^\dagger\mid0\rangle \end{equation}
$Q$ is the total momentum of the state, and both $Q$ and $k$ can be any vector in the first Brilluon zone. The Hamiltonian acts on the basis in the following way.
\begin{equation} H\mid Q,k\rangle = E_{k}\mid Q,k\rangle + \frac{U}{N_s}\sum_p\mid Q,p\rangle, \end{equation}
where $E_{k}=  (\epsilon_{Q-k}+\epsilon_k)$.
 The Hamiltonian conserves total momentum. Thus, we can diagonalize each total momentum sector independently. Each sector will have $N_s$ eigenstates, where $N_s$ is the size of the lattice. We now fix $Q$ and work in a particular total momentum sector. The basis states now depend on a single index $k$.
 The $E_{k}$ 's will in general be degenerate, and we take an $E$ with degeneracy $n$, i.e.  $ deg(E)=n$, corresponding to states 
 $\mid Q,k_1\rangle ... \mid Q,k_n\rangle$. From these we can make an $n-1$ dimensional degenerate eigenspace of the Hamiltonian with energy $E$ which we shall call $\mid\psi\rangle_{deg}$.

\begin{equation} \mid\psi\rangle_{deg} = \sum_{i=1}^n\alpha_i\mid Q,k_i\rangle\hspace{5mm}\sum_i\alpha_i=0 \label{degenerate} \end {equation}
\newline One can see that these are eigenstates with energy $E$ since potential energy term goes to zero due to the condition $\sum_i\alpha_i=0$ and the kinetic energy term gives $E$ times the state.
 Suppose there are $p$ unique values of $E$ in this total momentum sector.
\begin{equation} deg(E_1) + ... + deg(E_p)=N_s \end{equation}
 By forming states in the way described above, we can obtain $N_s - p$ eigenstates $\mid\psi\rangle_{deg}$ that are  independent of $U$.   We obtain the remaining non trivial (i.e. $U$ dependent) $p$ eigenstates by plugging the following state into the Hamiltonian.
 \beq
 \mid\psi_{Q}\rangle = \sum_k\Phi_{Q}(k) \mid Q,k\rangle  \;\;\; \mbox{ and} \;\;\; H \mid\psi_{Q}\rangle = \Lambda_{Q} \mid\psi_{Q}\rangle .
\eeq
Here we consider states with a fixed total momentum $Q$ since this object is conserved.
This yields the following results
\begin{equation} \Phi_{Q}(k) = \frac{1}{c_{Q}\sqrt{N_{s}}}\frac{1}{\Lambda_{Q}-E_k}, \;\;\;
 c_{Q} = (\frac{1}{N_s}\sum_k\frac{1}{(\Lambda_{Q}-E_k)^2})^\frac{1}{2},\;\;\; 
  \frac{U}{N_s}\sum_k\frac{1}{\Lambda_{Q}-E_k} = 1. \label{14} \end{equation}
   We can see explicitly from Eq.~(\ref{14}) that  $\langle\psi_{Q}\mid\psi\rangle_{deg}=0$ since basis states with equal $E$ have equal coefficients, and therefore the condition $\sum_{i}\alpha_{i}=0$  makes this state orthogonal to the degenerate manifold of states in Eq.~(\ref{degenerate}). There are $p-1$ solutions of Eq.~(\ref{14}) which lie in between the $p$ distinct $E$'s. The corresponding states are in the lower Hubbard band. The $ \mid\psi\rangle_{deg}$ found earlier also lie in the lower Hubbard band since these states are independent of $U$. There is one solution of \disp{14} for which $\Lambda_{Q}>E_{max}$ and is of order $U$ if $U>W$. The corresponding state lies in the upper Hubbard band. Thus for each fixed $Q$ sector, there is one state in the upper Hubbard band. We now consider the doublon state.
\begin{equation}  \mid\psi\rangle_d = c_{i\uparrow}^\dagger c_{i\downarrow}^\dagger\mid0\rangle = \frac{1}{N_s}\sum_{Q,k }e^{-iQ\cdot R_i}\mid Q,k\rangle \end{equation}

 We can rewrite  
\begin{equation}   \gamma(t_{r}) = \sum_Q\mid\langle\psi_{Q}\mid\psi\rangle_d\mid^2e^{-i \Lambda_{Q} t_{r}}  \label{ex-2}
\end{equation}
 where the $Q$ in the above sum stands for the $p$ states described by  Eq.~(\ref{14}) in the total momentum sector $Q$. Since $\langle\psi_d \ \mid\psi\rangle_{deg}=0$ we didn't have to take the degenerate states into account when calculating the correlation function.  Furthermore, we see that
\begin{equation}  \mid\langle\psi_{Q}\mid\psi\rangle_d\mid^2 = \frac{1}{N_sc_Q^2U^2} \end{equation}
 where $c_Q$  is from Eq.~(\ref{14}).  
\begin{equation} \gamma_U(t_{r}) = \sum_{Q\epsilon UHB}\frac{1}{N_sc_{Q}^2U^2} e^{-i\Lambda_{Q}t_{r}} \end{equation}
 In the above sum, each Q now represents only one state, since there is only one UHB state in each total momentum sector. We first evaluate this in one dimension, and then generalize to multiple dimensions. The sum can be turned into an integral.
\begin{equation} \gamma_U(t_{r}) = \frac{1}{\pi}\int_0^\pi\frac{1}{c_{Q}^2U^2} e^{-i\Lambda_{Q}t_{r}}dQ \end{equation}
 Converting Eq.~ (\ref{14})   into integrals, we find that 
\begin{equation} \Lambda_{Q} = (U^2+16t^2\cos^2\frac{Q}{2})^{\frac{1}{2}} \end{equation}
\begin{equation} c_{Q}^2 = \frac{1}{U^3}(U^2+16t^2\cos^2\frac{Q}{2})^{\frac{1}{2}} \end{equation}
 For $U > W$, we keep corrections of $O(\frac{t^2}{U^2})$ in $\Lambda_{Q}$ and drop all corrections in $c_{Q}^2$ , yielding
\begin{equation}  \gamma_U(t_{r}) \sim \frac{1}{\pi}\int_0^\pi e^{-iU(1+8\frac{t^2}{U^2}\cos^2\frac{Q}{2})t_{r}}dQ 
\lab{24}\end{equation}
\begin{equation}  \gamma_U(t_{r}) \sim e^{-i(U+4\frac{t^2}{U})t_{r}} J_0(\frac{4t^2}{U}t_{r}) \end{equation}
 In two dimensions, Eq.~(\ref{14}) becomes an elliptic integral so there is no closed form answer for the upper band eigenvalues in terms of elementary functions. However for  $U > W$, keeping corrections to the same order as we did in deriving Eq.~(\ref{24}), we can easily generalize to higher dimensions.
\begin{equation} \Lambda_{Q} = U(1+8\frac{t^2}{U^2}\Sigma_{i=1}^d\cos^2\frac{Q_i}{2}) \end{equation}
\begin{equation} c_{Q}^2 = \frac{1}{U^2} \end{equation}
\begin{equation}  \gamma_U(t_{r}) \sim e^{-i(U+4d\frac{t^2}{U})t_{r}} J_0^d(\frac{4t^2}{U}t_{r}) \end{equation}
 The other contribution to $\gamma(t_{r})$ is $\gamma_L(t_{r})$. However, from degenerate perturbation theory, we know that  provided $U > W$
 $\mid\langle\nu\mid\psi\rangle_d\mid^2$ is $O(\frac{t^2}{U^2})$ smaller for $\nu\epsilon LHB$ than it is for the upper Hubbard band. Hence, $\gamma_L(t_{r})$ is a small correction to $\gamma_U(t_{r})$.
\begin{equation} \gamma(t_{r}) \approx \gamma_U(t_{r}) \end{equation}
\begin{equation} \mid \gamma(t_{r})\mid^2 \approx J_0^{2d}(\frac{4t^2}{U}t_{r}) \end{equation}

 In conclusion, the doublon decay in the 2 particle Hubbard model in the regime $U>W$ is dominated by $\gamma_U$ with the much faster decaying $\gamma_L$ giving a small correction.  To a good approximation, the shape of the decay of $ \mid \gamma(t_{r})\mid^2$ is $J_0^{2d}(\frac{4t^2}{U}t_{r})$.


\begin{thebibliography}{235}
\bibitem{ECQL-linslope} In \refdisp{ECQL} Eq. (D4),  it is argued  that the real part of the self energy, in the range of frequencies $ W \ll \omega \ll U$ must have the behaviour $\Re e \Sigma(\vec{k}, \omega) = - \frac{n}{2-n} \omega + \mbox{const} $. Hence at a density $n=.0414$ we expect  $\Re e \Sigma(\vec{k}, \omega) = - .0212 \omega + \mbox{const}$, as found numerically.
\bibitem{derivation1}
This is easy to see from the definition of the scattering amplitude and the its result in the ladder scheme:
\barray
G^{II}(Q,p_1,p_2) &=& G(Q/2+p_1)G(Q/2-p_1)-G(Q/2+p_1)G(Q/2-p_1)\Gamma(Q)G(Q/2+p_2)G(Q/2-p_2)\nonumber\\
\gamma(Q) &=& \sum_{p_1,p_2}G^{II}(Q,p_1,p_2)=\frac{\Pi(Q)}{1+U\Pi(Q)}, \nn
\earray
where we used \disp{eq1}.
The spectral density of $\gamma$ is now easy to find 
\beq
\rho_{\gamma}=\frac{-\rho_{\Pi}}{(1+U Re\Pi)^2+(\pi U \rho_{\Pi})^2}, \nn
\eeq
leading to the quoted result.
\end{thebibliography}
\end{document}